\documentclass{sig-alternate-10pt}
\pdfoutput=1

\usepackage{times}
\newcommand{\subparagraph}{}
\usepackage{amsmath}
\usepackage{amsfonts}
\usepackage{amssymb}
\usepackage{xspace}
\usepackage{xparse}
\usepackage[dvipsnames]{xcolor}
\usepackage[font=bf]{caption}
\usepackage[caption=false]{subfig}
\usepackage{wrapfig}
\usepackage{tikz}
\usepackage{graphicx}
\definecolor{shadecolor}{gray}{0.9}
\usepackage{framed}
\usepackage[breaklinks=true]{hyperref}

\usepackage{booktabs}
\usepackage{multirow}
\usepackage{listings}
\usepackage{tabularx}
\usepackage{tabulary}

\newcommand{\secref}[1]{\S\ref{#1}\xspace}

\newcommand{\figref}[1]{Fig.~\ref{#1}\xspace}
\newcommand{\tabref}[1]{Table~\ref{#1}\xspace}

\newcommand{\CPLEX}{\texttt{CPLEX}\xspace}
\newcommand{\Gurobi}{\texttt{Gurobi}\xspace}
\newcommand{\OpenDaylight}{\texttt{OpenDaylight}\xspace}

\newcommand{\Python}{\texttt{Python}\xspace}

\newcommand{\nodeSet}{\texttt{nodes}\xspace}
\newcommand{\nodeSubset}{\ensuremath{N}\xspace}

\newcommand{\edgeSet}{\texttt{links}\xspace}
\newcommand{\node}{\ensuremath{v}\xspace}
\newcommand{\edge}{\ensuremath{l}\xspace}
\newcommand{\commodity}{\ensuremath{c}\xspace}
\newcommand{\commoditySet}{\texttt{classes}\xspace}
\newcommand{\commoditySubset}{\ensuremath{C}\xspace}

\newcommand{\Path}{\ensuremath{p}\xspace}

\newcommand{\pathSet}[1]{\ensuremath{\texttt{paths}({#1})}\xspace}

\newcommand{\nodeLoad}[2]{\ensuremath{\mathit{load}_{#1}^{#2}}}
\newcommand{\edgeLoad}[2]{\ensuremath{\mathit{load}_{#1}^{#2}}}
\newcommand{\nodeCapacityVar}[2]{\ensuremath{\mathit{capvar}_{#1}^{#2}}\xspace}
\newcommand{\flowFrac}[2]{\ensuremath{x_{{#1},{#2}}}\xspace}

\newcommand{\enabled}[1]{\ensuremath{b_{#1}}\xspace}
\newcommand{\FrName}{SOL\xspace}
\newcommand{\Name}{\FrName}

\newcommand{\allocation}[1]{\ensuremath{a_{#1}}}
\newcommand{\true}{\texttt{True}\xspace}
\newcommand{\false}{\texttt{False}\xspace}

\newcommand{\resource}{\ensuremath{r}\xspace}
\newcommand{\edgeResource}{\ensuremath{\resource}\xspace}
\newcommand{\nodeResource}{\ensuremath{\resource}\xspace}
\newcommand{\edgeCapacity}{\ensuremath{lnCap}\xspace}
\newcommand{\nodeCapacity}{\ensuremath{ndCap}\xspace}
\newcommand{\totalCapacity}{\ensuremath{totCap}\xspace}

\newcommand{\nodeBudget}{\ensuremath{k}\xspace}
\newcommand{\pathPredicate}{\texttt{predicate}\xspace}
\newcommand{\pathPruneStrategy}{\texttt{selectStrategy}\xspace}
\newcommand{\pathPruneStrategyShortest}{\texttt{shortest}\xspace}
\newcommand{\pathPruneStrategyRandom}{\texttt{random}\xspace}
\newcommand{\pathPruneNmbr}{\texttt{selectNumber}\xspace}

\newcommand{\ExampleNode}[1]{\ensuremath{\text{N{#1}}}\xspace}
\newcommand{\ExampleCommodity}[1]{\ensuremath{\text{C{#1}}}\xspace}

\newcommand{\labelAllocateFlow}{\texttt{addAllocateFlowConstraint}\xspace}
\newcommand{\labelRouteAllFlow}{\texttt{addRouteAllConstraint}\xspace}
\newcommand{\labelLinkCon}{\texttt{addLinkCapacityConstraint}\xspace}
\newcommand{\labelNodeCon}{\texttt{addNodeCapacityConstraint}\xspace}
\newcommand{\labelNodeConPerPath}{\texttt{addNodeCapacityPerPathConstraint}\xspace}
\newcommand{\labelCapacityBudget}{\texttt{addCapacityBudgetConstraint}\xspace}
\newcommand{\labelAllNodes}{\texttt{addRequireAllNodesConstraint}\xspace}
\newcommand{\labelSomeNodes}{\texttt{addRequireSomeNodesConstraint}\xspace}
\newcommand{\labelAllEdges}{\texttt{addRequireAllEdgesConstraint}\xspace}
\newcommand{\labelEnforcePathDisable}{\texttt{addPathDisableConstraint}\xspace}
\newcommand{\labelOnePath}{\texttt{addEnforceSinglePath}\xspace}
\newcommand{\labelNodeBudget}{\texttt{addBudgetConstraint}\xspace}
   
\newcommand{\labelSetPredefObj}{\texttt{setPredefinedObjective}\xspace}

\newcommand{\labelMaxNodeLoad}{\texttt{minMaxNodeLoad}\xspace}

\newcommand{\labelMaxLinkLoad}{\texttt{minMaxLinkLoad}\xspace}
\newcommand{\labelTotalFlow}{\texttt{maxAllFlow}\xspace}

\newcommand{\labelRoutingCost}{\texttt{minRoutingCost}\xspace}

\newcommand{\edgeCapFunc}{\texttt{linkCapFn}\xspace}
\newcommand{\nodeCapFunc}{\texttt{nodeCapFn}\xspace}
\newcommand{\nodeBudgetFunc}{\texttt{nodeBudgetFn}\xspace}
\newcommand{\routingCostFunc}{\texttt{routingCostFn}\xspace}

\newcommand{\xp}{\texttt{xp}\xspace}
\newcommand{\al}{\texttt{al}\xspace}
\newcommand{\bn}{\texttt{bn}\xspace}
\newcommand{\be}{\texttt{be}\xspace}
\newcommand{\bp}{\texttt{bp}\xspace}

\newcommand{\el}{\texttt{el}\xspace}
\newcommand{\nl}{\texttt{nl}\xspace}
\newcommand{\nc}{\texttt{nc}\xspace}
\newcommand{\defVar}{\texttt{defineVar}\xspace}
\newcommand{\setObj}{\texttt{setObjective}\xspace}
\newcommand{\lowerBound}{\ensuremath{\mathit{lb}}\xspace}
\newcommand{\upperBound}{\ensuremath{\mathit{ub}}\xspace}
\newcommand{\name}{\ensuremath{\mathit{name}}\xspace}
\newcommand{\coeffs}{\ensuremath{\mathit{coeffs}}\xspace}
\newcommand{\None}{\texttt{TBA}\xspace}
\newcommand{\optDirection}{\ensuremath{\mathit{dir}}\xspace}

\newcommand{\newOpt}{\texttt{getOptimization}\xspace}

\newcommand{\optObj}{\texttt{opt}\xspace}

\newcommand{\comment}[1]{}
\newcommand{\ignore}[1]{}

\definecolor{darkgreen}{HTML}{00AB00}

\newcounter{packednmbr}

\newenvironment{packeditemize}{
\begin{list}{$\bullet$}{
    \setlength{\itemsep}{0pt}
    \addtolength{\labelwidth}{10pt}
    \setlength{\leftmargin}{12pt}
    \setlength{\listparindent}{\parindent}
    \setlength{\parsep}{2pt}
    \setlength{\topsep}{0pt}}}
    {\end{list}
}

\newcommand{\tightcaption}[1]{\caption{\em #1}\vspace{-0.3cm}}

\newcommand{\mypara}[1]{\smallskip\noindent{\bf {#1}:}~}

\graphicspath{{./figures/}}
\hypersetup{
pdftitle=\FrName,
pdfsubject=\FrName,
colorlinks=true,
linkcolor=Sepia,
urlcolor=NavyBlue,
citecolor=Brown
}

\date{}
\title{
Accelerating the Development of Software-Defined Network  Optimization Applications Using \Name
}
\numberofauthors{3}
\author{
    \alignauthor Victor Heorhiadi\\
    \affaddr{UNC Chapel Hill}\\
    \email{victor@cs.unc.edu}
    \alignauthor Michael K. Reiter\\
    \affaddr{UNC Chapel Hill}\\
    \email{reiter@cs.unc.edu}
    \alignauthor Vyas Sekar\\
    \affaddr{Carnegie Mellon University}\\
    \email{vsekar@andrew.cmu.edu}
}

\begin{document}
\maketitle

\begin{abstract}
Software-defined networking (SDN) can enable diverse network management
applications such as traffic engineering, service chaining, network function
outsourcing, and topology reconfiguration. Realizing the benefits of SDN for
these applications, however, entails addressing complex  network optimizations
that are central to these problems. Unfortunately, such optimization problems
require  significant manual effort and expertise to express and non-trivial
computation and/or carefully crafted heuristics to solve. Our vision is to
simplify the deployment of  SDN applications using {\em general} high-level
abstractions for capturing  optimization requirements from which we can {\em
efficiently} generate optimal solutions. To this end, we present  \FrName, a
framework that demonstrates that it is indeed possible to simultaneously achieve
generality and efficiency. The insight underlying \FrName is that SDN
applications can be recast within a unifying {\em path-based} optimization
abstraction, from which it efficiently generates near-optimal solutions, and
device configurations to implement those solutions.

We illustrate the generality of \FrName by prototyping diverse and new
applications. We show that \Name simplifies the development of SDN-based
network optimization applications and provides comparable or better scalability
than custom optimization solutions.
\end{abstract}

\section{Introduction}
\label{sec:intro}

A key motivation for software-defined networking (SDN) is that it is
an  enabler for  network management applications that would
otherwise be difficult to realize using existing control-plane
mechanisms.  In particular, the consolidation of control
logic enables operators to systematically implement network
configuration for traffic engineering (e.g.,~\cite{measurouting}),
service chaining (e.g.,~\cite{simple}), energy efficiency
(e.g.,~\cite{elastictree}), network function virtualization (NFV)
(e.g.,~\cite{stratos}), and cloud offloading (e.g.,~\cite{aplomb}),
among others.

While this body of work has been instrumental in demonstrating the
potential benefits of SDN, realizing these benefits requires
significant effort.  In particular, at the core of many SDN
applications are custom optimization problems to tackle various
constraints and requirements that arise in practice. (We elaborate on
this in \secref{sec:background}.)  For instance, an SDN application
might need to account for limited TCAM, link capacities, or middlebox
capacities, among other considerations.  Developing such optimization
formulations involves a non-trivial learning curve, a careful
understanding of both theoretical and practical issues, and
considerable manual effort.  Furthermore, when the resulting
optimization problems are intractable to solve with state-of-the-art 
 solvers (e.g., \CPLEX or \Gurobi), heuristic algorithms must be
crafted to ensure that new device configurations can be generated on
the timescales demanded by the application as relevant inputs (e.g.,
traffic matrix entries) change (e.g.,~\cite{elastictree,panopticon}).
Finally, without a common framework for representing network
optimization tasks, it is difficult to reuse key ideas from one
application or to combine useful features from multiple applications
into a custom new application.  As such, many efforts effectively
reinvent common building blocks, e.g., ensuring that the rules output
by the optimizations can fit inside the TCAM, or generating
volume-aware load-balancing rules that also maintain flow affinity.

\begin{figure}[t]
\includegraphics[width=\columnwidth]{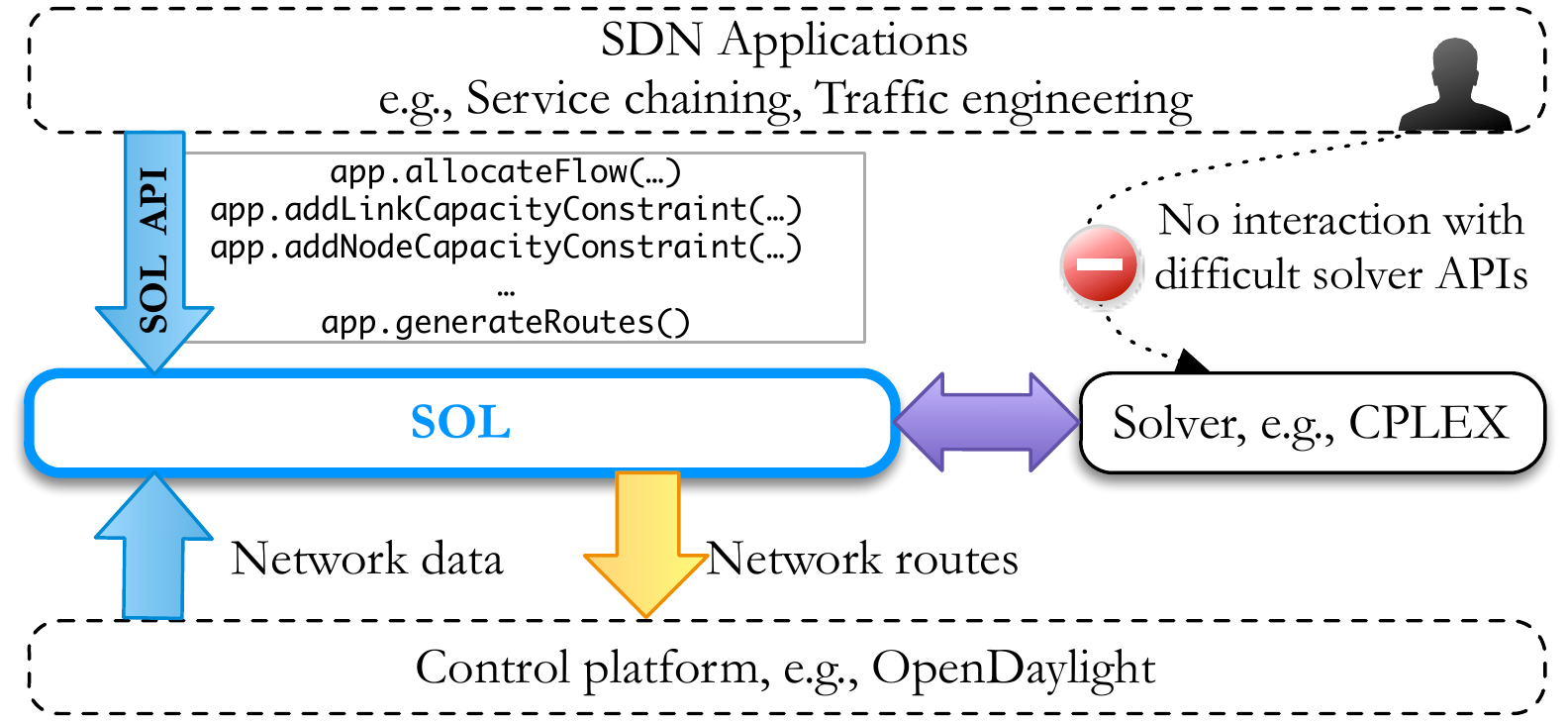}
\tightcaption{Overview of \FrName: Developers use the \FrName
  high-level APIs to specify optimization goals and
  constraints. \FrName generates near-optimal solutions and produces
  device configurations that are input to the SDN control platform.}
\label{fig:overview}
\end{figure}

Our goal in this work is to raise the level of abstraction for writing
SDN applications by providing a flexible optimization framework to
reduce the development time to build such applications atop SDN.  In
doing so, we seek to minimize the ``pain
point'' for each network optimization application. 
 To this end, here we introduce
\FrName, a framework that enables SDN application developers to
express new application goals and constraints in a high-level
language, and then generates compliant configurations that can be
deployed to SDN control platforms directly (see
\figref{fig:overview}).  Conceptually, \FrName can be viewed as an
intermediate layer that sits between the SDN optimization applications
and the actual control platform. Application developers who want to
develop new network optimization capabilities write their optimization
formulations using the \FrName API.

There are two requirements for a framework like \FrName:
\begin{packeditemize}
\item {\bf Generality:}  First, we want \FrName to
be \textit{general}; i.e.,  capable of expressing the 
 requirements  of optimization problems across the full spectrum of 
 SDN applications (e.g., traffic engineering, policy
steering, load balancing, and topology management). This
 entails abstractions
that bridge the gap between SDN practitioners and mathematical
optimization tools.  

\item {\bf Efficiency:} Second, \FrName should be \textit{efficient},
generating compliant and optimal (or near-optimal) 
configurations on a timescale that is responsive to application needs
as new inputs (e.g., traffic-matrix changes) become known.
\end{packeditemize}

Given the diversity of the application requirements and the trajectory of prior
work
(e.g.,~\cite{simple,b4,swan,elastictree,panopticon,stratos,jrex,response,measurouting,snips}),
 generality and efficiency appear individually difficult, let alone
achieving both simultaneously.  Our contribution in this paper is (perhaps
surprisingly) to show that this is indeed possible, and that generality does
not need to come at the expense of efficiency.

Our key insight to achieve {\em generality} is that SDN applications
and their associated optimization problems can be expressed as {\em
  path-based} optimization problems.  It is very natural for
application developers to think in terms of the paths that packets
traverse through the network.  Moreover, it becomes easy to express
key policy requirements in terms of those paths.  For example, we can
specify service chaining requirements (e.g., each path includes a
firewall and intrusion-detection system (IDS), in that order) or
redundancy (e.g., each includes two intrusion-prevention systems
(IPS), in case one fails open).  Finally, it is easy to model device
resource consumption, such as routing table and TCAM space, based on
the traffic carried on paths that traverse that device.

The natural question then is if this generality sacrifices efficiency.
Indeed, if implemented naively, optimization problems expressed over the paths
that traffic might travel will  introduce efficiency challenges since the
number of paths  grows exponentially with the network size.  Our insight, however,
that it is not necessary to consider them all.  Specifically, we show that by
combining infrequent, offline preprocessing with simple, online {\em
path-selection algorithms} (e.g., shortest paths or random paths), \FrName
achieves near-optimal solutions in practice for all applications we considered.
Moreover, \Name is typically far more efficient than solving the optimization
problems originally used to express these applications' requirements.

We have implemented \FrName as a Python-based library that interfaces
with \OpenDaylight (\secref{sec:implementation}).  We have also
prototyped numerous SDN applications in \FrName, including
SIMPLE~\cite{simple}, ElasticTree~\cite{elastictree}, 
Panopticon~\cite{panopticon}, and others of our own design
(\secref{sec:examples}).  We are planning an open-source release of
\FrName, and these and other applications coded in \FrName, in the
near future.  Our evaluations 
on a range of topologies show that: 1) \FrName outperforms several
of these applications' native optimization algorithms; 
2) \FrName scales better than other network management tools like
Merlin~\cite{merlin}; and 3) \FrName substantially reduces the effort
required (e.g., in terms of lines of code) for implementing new SDN
applications.

\section{Background and Motivation}
\label{sec:background}

In this section, we describe representative network  applications that could
benefit from a framework such as \FrName.  We  highlight  the need for careful
formulation and algorithm development involved in prior efforts, as well  the
diversity of requirements they entail.

\subsection{Traffic engineering} 
\label{sec2:te}

\begin{figure}[h]
\vspace{-.1in}
\centering
\includegraphics[width=.45\textwidth]{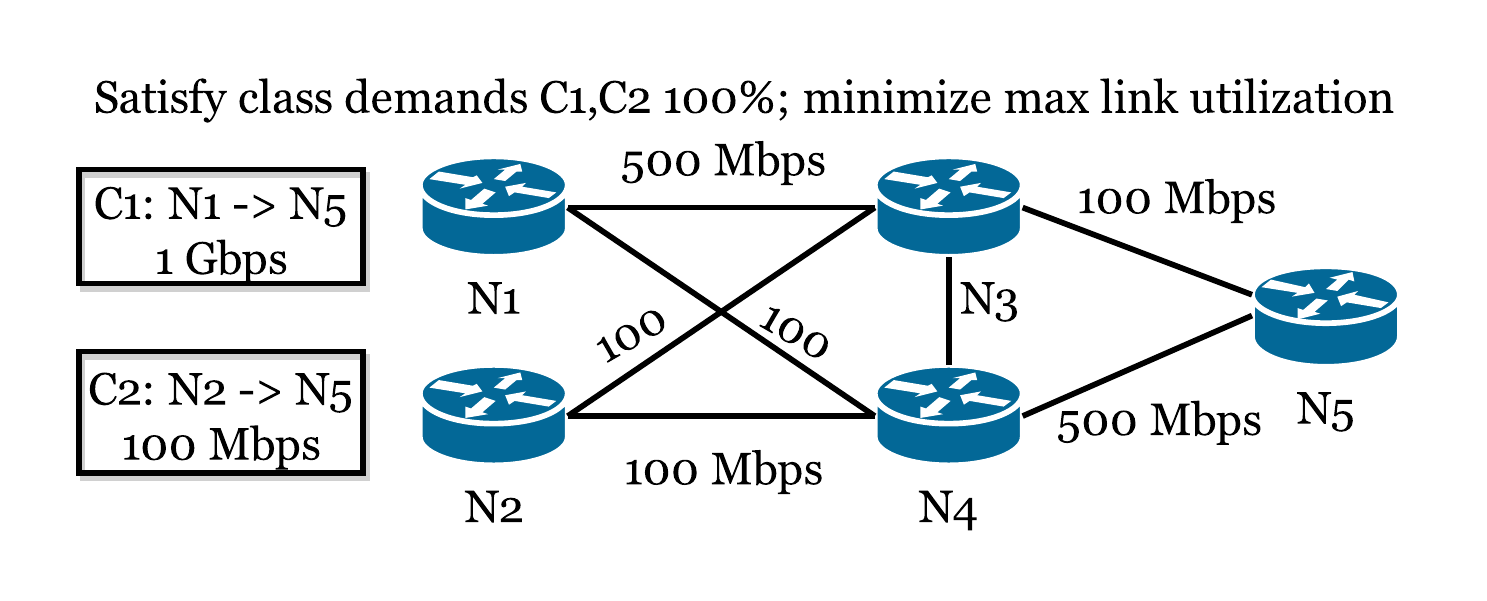}
\vspace{-.1in}
\tightcaption{Traffic engineering applications}\label{fig:bg:te}
\end{figure}

Traffic engineering (TE) is a canonical network management application that is
one of the early driving applications for SDN~\cite{b4,swan}.
\figref{fig:bg:te} shows an example where classes \ExampleCommodity{1} and
\ExampleCommodity{2} need to be routed completely while minimizing the load on
the most heavily loaded link.  At a high-level, the inputs to a TE application
are the traffic demands (e.g., the traffic matrix between WAN sites) and a
specification of the traffic {\em classes} and priorities. The TE application
takes into account the network topology and link capacities to determine how to
route each class to achieve network-wide objectives; e.g., simple objectives
like minimize maximum congestion~\cite{thorup} or  weighted max-min
fairness~\cite{b4,swan}.  

\mypara{Challenges}
While simple goals like minimizing link congestion can be compactly
represented and solved via max-flow formulations~\cite{networkflowbook}, the
expressivity and efficiency quickly breaks down for more complex
objectives such as weighted max-min fairness~\cite{b4,swan}. In fact,
a significant part of the technical contribution in SWAN is the
development of an algorithm for max-min
fairness~\cite{swan}. Similarly, B4~\cite{b4} has implemented several
theoretical optimizations for scalability~\cite{danna2012practical}. When
max-flow like formulations fail, designers invariably revert to
``low-level'' first-principles techniques such as linear programs (LP)
or combinatorial algorithms.  Neither is desirable; using and
debugging with LP solvers is painful as they expose a very low-level
interface, and combinatorial algorithms require significant
theoretical expertise.  Finally, translating the algorithm output into
actual routing rules requires care to realize the optimization
benefits; e.g., the rules must be volume-aware to realize the benefits
of the TE policy~\cite{openflowlb}.

\subsection{Service chaining}
\label{sec2:sc}
\begin{figure}[h]
\vspace{-.05in}
\centering
\includegraphics[width=.45\textwidth]{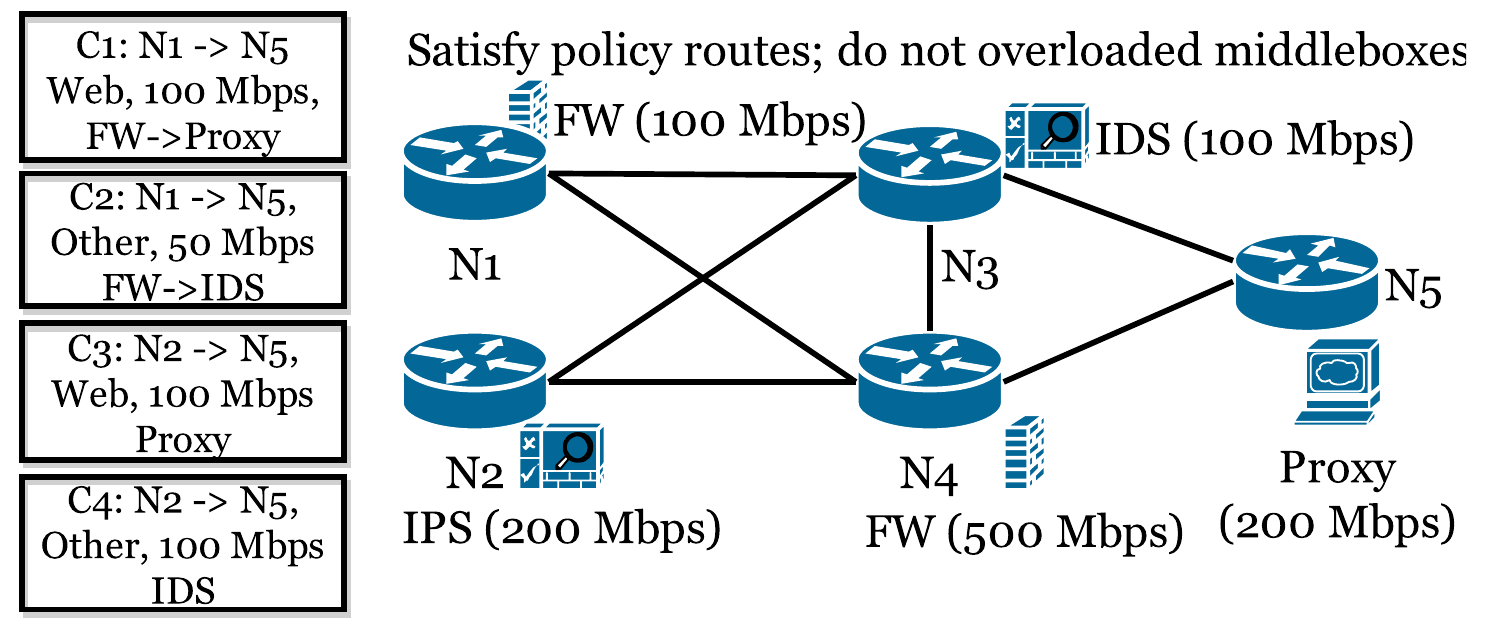}
\vspace{-.1in}
\tightcaption{Service chaining applications}\label{fig:bg:service}
\end{figure}

Networks today rely on a wide variety of third-party appliances or middleboxes
for performance, security, and policy compliance capabilities such as intrusion
detection and prevention systems, firewalls, application-layer proxies, and load
balancers (e.g.,~\cite{aplomb}).  The goal of service chaining is to ensure that
each class of traffic is routed through the desired sequence of network
functions to meet the policy requirements.  For example, in
\figref{fig:bg:service}, class \ExampleCommodity{1} is required to traverse a
firewall and proxy, in that order. Such policy routing rules must be suitably
encoded within the available TCAM on SDN switches~\cite{simple}.  Since some
middleboxes implement expensive computations, they can get easily overloaded and
thus operators would like to balance the load on these
appliances~\cite{simple,opennf}.  Thus, the key inputs to such applications are
the different traffic policy classes, their service chaining requirements, and
the available middlebox resources in the network. The application then sets up
the forwarding rules such that the policy requirements for service chaining are
met while respecting the switch TCAM and middlebox capacities.  Finally, as a
practical constraint, many of these middleboxes are stateful, so the
load-balancing rules that are installed must ensure flow affinity.

\mypara{Challenges}
Service chaining introduces more complex requirements when compared to
TE applications.  First, modeling the consumption of switch TCAM
introduces ``discreteness'' into the optimization problem, which
impacts scalability~\cite{simple}.  Second, expressing such service
processing requirements falls outside the scope of existing network
flow abstractions~\cite{jrex}.  Third, service chaining also
highlights the complexity of combining different network requirements;
e.g., reasoning about the interaction between the load balancing
algorithm and the switch TCAM constraints is quite non-trivial and
introduces circular dependencies~\cite{softcell}.  Existing service chaining
efforts have to develop custom heuristics~\cite{cao2014traffic} or need new
theoretical extensions~\cite{jrex}.  Finally, ensuring requirements
like flow affinity is tricky due to routing asymmetry~\cite{conext12} or due
to potential rule conflicts~\cite{snips} and thus require non-trivial care.

\subsection{Flexible topology management}
\label{sec2:topology}
 
\begin{figure}[h]
\vspace{-.1in}
\centering
  \includegraphics[width=.45\textwidth]{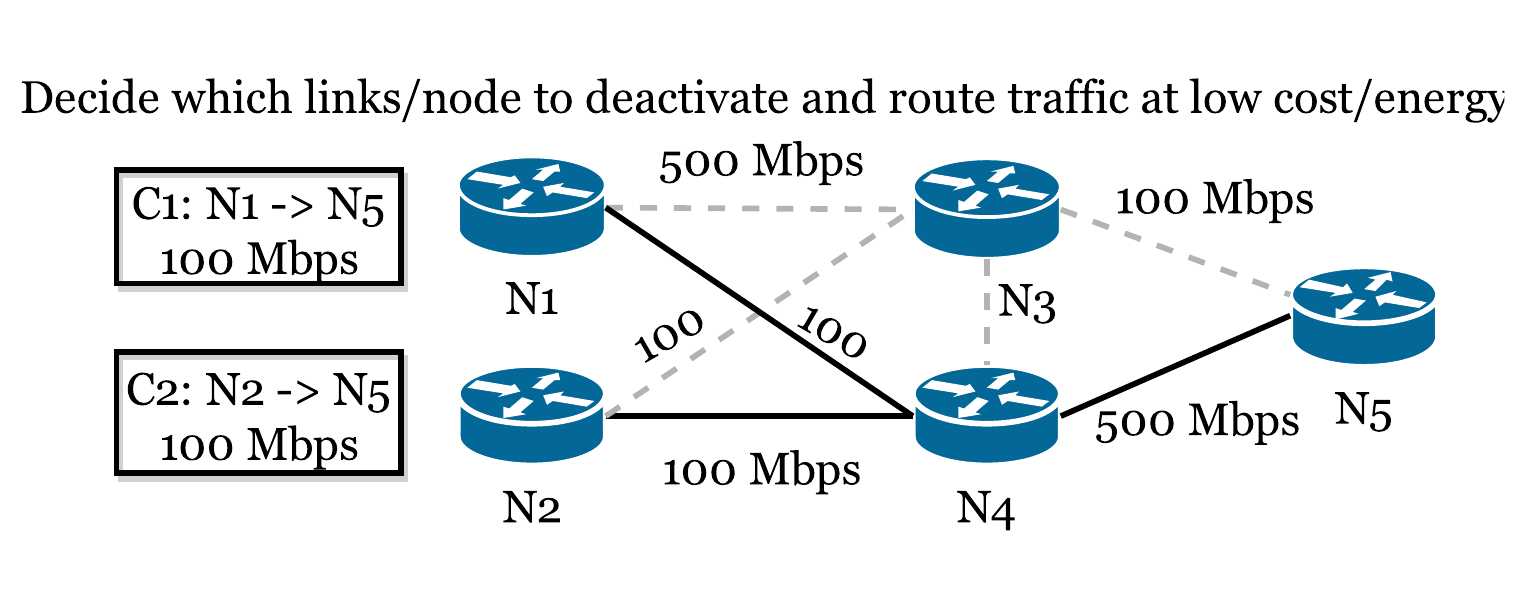}
\vspace{-.1in}
\tightcaption{Topology reconfiguration applications}\label{fig:bg:topology}
\end{figure}

SDN can enable new topology modification capabilities that would
otherwise be difficult, if not impossible, to implement with existing
control plane techniques.  For instance, prior work such as
ElasticTree~\cite{elastictree} and Response~\cite{response} has used
SDN to dynamically switch on/off network links and nodes to make
datacenters more energy efficient.  In \figref{fig:bg:topology}, these
applications might shut down node \ExampleNode{3} during periods of
low utilization, if routing classes \ExampleCommodity{1} and
\ExampleCommodity{2} via \ExampleNode{4} does not significantly impact
end-to-end performance.  Topology reconfiguration is especially
feasible in rich datacenter topologies with multiple paths between
every source and destination.  Such applications take as input the
demand matrix (similar to the TE task) and then compute both which
nodes and links should be active and traffic-engineered routes to
ensure some performance service-level agreements.

\mypara{Challenges} The on-off requirement on the switches/links once
again introduces discrete constraints, yielding integer-linear
optimizations that are theoretically intractable and difficult to
directly express using common max-flow like abstractions.  As prior
work shows, solving such a problem requires significant computation
even on small topologies and thus we may have to develop new heuristic
solving strategies; e.g., ElasticTree uses a greedy bin-packing
algorithm~\cite{elastictree}.

\subsection{Network function virtualization} 
\label{sec2:nfv}

\begin{figure}[h]
\includegraphics[width=.45\textwidth]{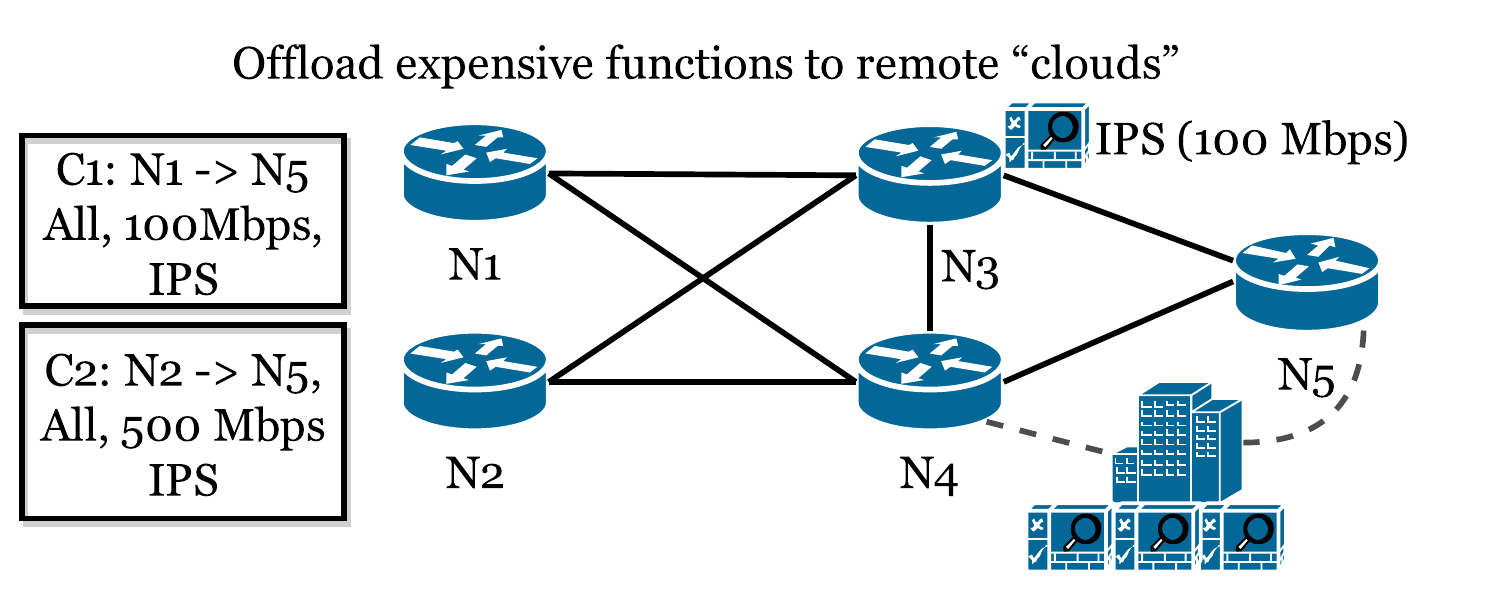}
\tightcaption{Offloading network functions}\label{fig:bg:offload}
\end{figure}

One of the key advantages of SDN is that it decouples the
logical specification of network requirements from their physical realization.
Prior work has leveraged SDN capabilities to offload or outsource such
expensive functions to leverage  clusters or
clouds~\cite{aplomb,gibb2012outsourcing,measurouting}.  This is especially
useful in the context of complex and expensive deep-packet-inspection services
that networks need today~\cite{snips}.   The key decision here is to decide how
much of the processing on each path to offload to the remote datacenter ---
e.g., in \figref{fig:bg:offload}, how much of class \ExampleCommodity{1}
traffic should be routed to the datacenter between \ExampleNode{4} and
\ExampleNode{5} for IPS processing, versus processing it at \ExampleNode{3}.
Offloading can increase user-perceived latency and impose additional load on
network links. Moreover, some active functions (e.g., WAN optimizers or IPS) may
induce changes to the observed traffic volumes due to their actions. Thus,
optimizing such offloading must take into account the congestion that might be
introduced, as well as latency impact and any traffic volume changes induced by
such outsourced functions.  Generalizing this further, we can also envision
novel {\em elastic scaling} opportunities where the number of middleboxes
required can even be scaled on demand~\cite{stratos,etsinfv,dpiserviceconext14}.

\mypara{Challenges}
These kind of offloading and elastic scaling opportunities introduce
new dimensions to optimization that are painful to capture. For
instance, with offloading we are effectively rerouting the traffic and
thus we need to carefully model the impact on TE objectives and link
loads and on downstream nodes. If done naively, this can introduce
non-linear dependencies since the actions of downstream nodes depend
on control decisions made upstream. To complicate matters further,
many of these network functions may actively change the traffic
volumes (e.g., compression for redundancy elimination or drops by
IPS).  Once again, if not done carefully,
this introduces non-linear dependencies in the optimization. Finally,
elastic scaling once again introduces a discrete aspect to the problem
similar to the topology modification application above and further
makes the problem intractable.

\subsection{Motivation for \FrName}

Drawing on the  above discussion (and our own experience) we summarize a few
key considerations: 

\begin{packeditemize}
\item Network applications have diverse and complex optimization
  requirements; e.g., service chaining requires us to reason about
  ``valid'' paths and topology modification needs to enable/disable
  nodes.

\item Designers of these applications have to spend significant effort
  in expressing and debugging these problems using low-level
  optimization libraries.

\item It can take non-trivial expertise to ensure that the
  problems can be solved fast enough to be relevant for operational
  timescales, e.g., recomputing TE every few minutes or periodically
  solving the large ILPs characteristic of topology reconfiguration
  applications (e.g.,~\cite{elastictree}).
\end{packeditemize}

We argue that if we had better general abstractions for expressing these
applications and tools for efficiently solving these problems, the time and
effort spent in these aforementioned efforts (and future SDN applications) can
be dramatically reduced. Based on the diversity of the requirements we see above
and the non-trivial effort\footnote{While it is hard to directly estimate the
number of human-hours spent, using the real estate in these papers devoted to
the optimization component as a rough proxy suggests non-trivial effort.} that
these prior efforts have spent in optimization algorithm development, such a
goal may seem elusive. Fortunately, as we show in the rest of the paper, this
goal can be realized.

\section{\FrName Overview}
\label{sec:overview}

Our overarching vision in developing \FrName is to raise the level of
abstraction in developing new SDN applications and specifically to
eliminate some of the black art in developing SDN-based optimizations,
making them more accessible for deployment by network managers.
\FrName takes in as inputs the network topology and traffic patterns
and the optimization requirements specified by clients in the \FrName
API. It will then translate these into low-level constraints in the
language of existing optimization solvers such as \CPLEX or \Gurobi.
Finally, \FrName interfaces with existing SDN control platforms such
as \OpenDaylight to install the desired forwarding rules on the SDN
switches and also install other middlebox-specific configuration
parameters. \FrName does not require modifications to the existing
control or data plane components of the network.  Our vision for
\FrName stands in stark contrast to the state of affairs today, in
which a developer faces programming a new SDN optimization either
directly for a generic and low-level optimization solver such as
\CPLEX or using a typically heuristic algorithm designed by hand,
after which she must translate the decision variables of the
optimization to device configurations.

\mypara{Path abstraction}
For \FrName to be useful and robust, we need a unifying abstraction
that can capture the requirements of diverse classes of SDN network
optimization applications described in the previous section.  \FrName
is built using \textit{paths} through a network as a core abstraction
for expressing network optimization problems.  This is contrary to how
many optimizations are formulated in the literature --- using a more
standard edge-centric approach~\cite{networkflowbook}.  In our
experience, however, an edge-centric approach forces complexity when
presented with additional requirements, especially ones that attempt
to capture path properties~\cite{panopticon,elastictree}.

In contrast, path-based formulations capture these requirements more
naturally.  For instance, much of the complexity in the modeling for
policy steering or offloading applications from
\secref{sec:background} is in capturing the path properties that need
to be satisfied. With a path-based abstraction, we can simply define
predicates that specify valid paths --- e.g., those that include
certain waypoints or that avoid a certain node (to anticipate that
node's failure).  In addition, we can model path-based resource use
with ease. For example, usage of TCAM space in a switch corresponds to
a traffic-carrying path traversing that switch (and thus a rule to
accommodate that path).  Without the path abstraction, modeling such
constraints is difficult (cf.,~\cite{simple}).  Finally, expressing
constraints on nodes and edges requires little change and does not
introduce increased difficulty.

\mypara{Tractability}
In a pure flow-routing scenario, an edge-based formulation admits
simple algorithms that guarantee polynomial-time execution.
Path-based formulations, on the other hand, are often dismissed
because of their inefficient appearance --- after all, in the worst
case, number of paths in the network is exponential with respect to
network size --- or due to the complexity of algorithms to solve path
based formulations (column-generation, decompositions,
etc.~\cite{networkflowbook}).  However, in many practical scenarios,
the number of valid paths (as defined by the application) is likely to
be significantly smaller. Furthermore, multipath routing can provide
only so much network diversity before its value
diminishes~\cite{liu2014multipath}.  So, the set of paths considered
need not be large.

\begin{figure}[th]
\centering
\includegraphics[width=0.8\columnwidth]{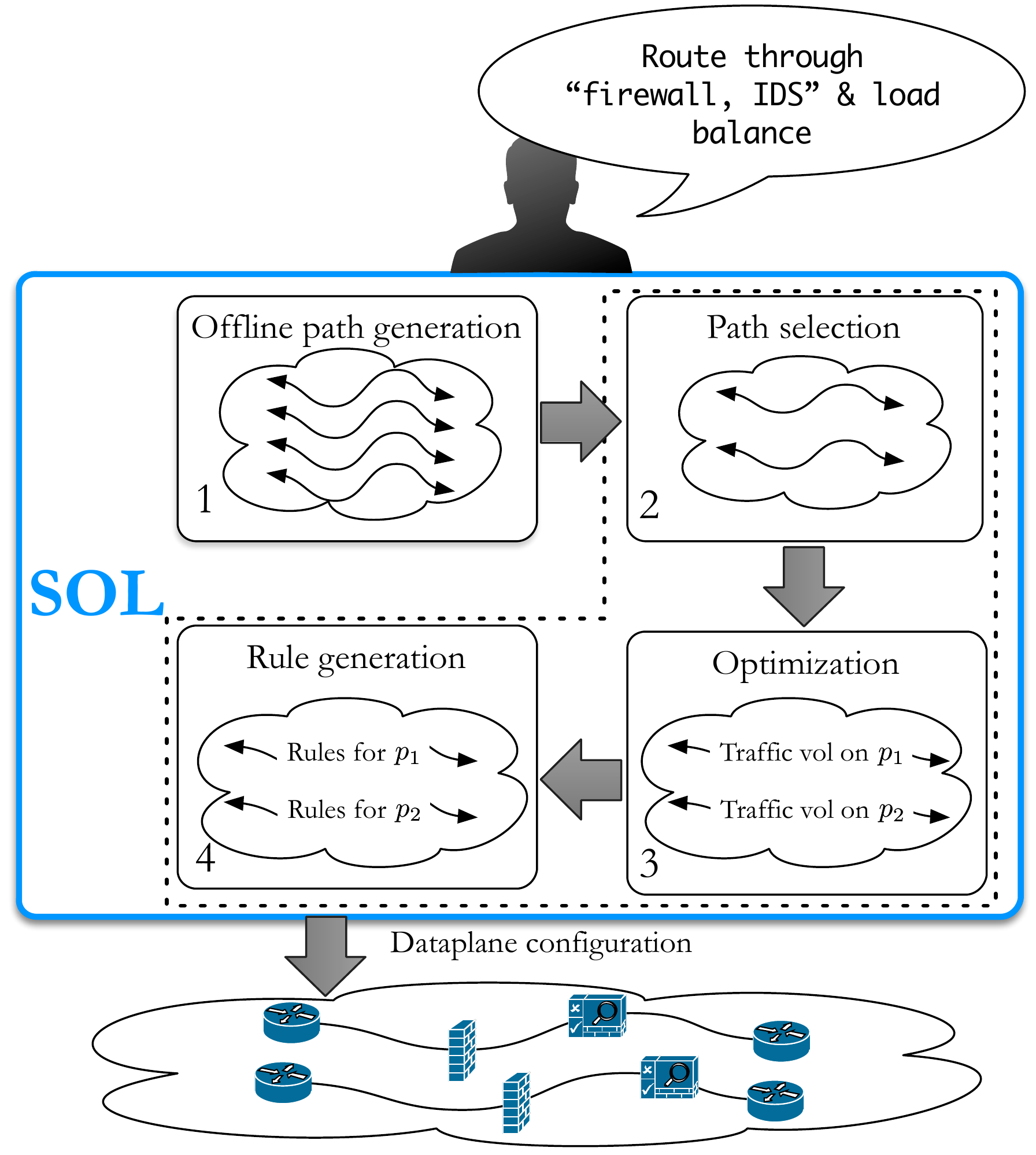}
\tightcaption{\FrName architecture, overview of the workflow}
\label{fig:arch}
\end{figure}

\FrName leverages an off-line path generation step to determine valid
paths (step 1 of \figref{fig:arch}). Since for most applications, the
set of valid paths is fairly static and does not need to be recomputed
every time the optimization is run, this step is infrequent.  Next,
\FrName \emph{selects} a subset of these paths (step 2) using a
selection strategy (see \secref{sec:path}) and runs the optimization
using only the selected paths as those available (step 3), to ensure
that the optimization completes quickly.  We show in \secref{sec:eval}
that this strategy still permits inclusion of sufficiently many paths
for the optimization to converge on a (near) optimal value.  So, while
the efficiency of path-based optimization is a valid theoretical
concern, in practice we show that there are practical heuristics to
address this issue.

\mypara{Generating device configurations}
\FrName directly translates the decision variables of a \FrName
optimization to network device configurations to, for example,
implement appropriate flow routing (step 4 of \figref{fig:arch}).  The
algorithm utilized in \FrName to perform this translation is based on
that in previous work~\cite{openflowlb,snips}, but is critically
enabled by the path-based formulations in \FrName{} --- one more
advantage of the path abstraction that we employ.  In contrast, the
original optimization formulations of some applications we consider
require the development of an additional custom algorithm to map their
decision variables to device configurations.

\section{\Name Detailed Design}
\label{sec:interface}

In this section, we present the detailed design of \Name.  We focus on  the
high-level API that the SDN application developer would  use to express
applications via \Name.    For completeness,  however, we present the  formal
basis that \Name uses {\em internally} for each such template.

Note, however, that the developer ``thinks'' in terms of the high-level API
rather than low-level details of dealing with the solver-level variables, how
paths are identified, etc.  A developer begins a \Name-based optimization by
instantiating a new \optObj object via the \newOpt function and then proceeds to
build the optimization using the {\em constraint templates}, which we explain
below.\footnote{Due to space limits we focus on salient aspects of the API and
refer readers to our (anonymized) manual~\cite{manual}.}

\subsection{Preliminaries}

\label{sec:interface:input}
\begin{figure}[t]
    \begin{minipage}[t]{\columnwidth}
        \raggedright
        \begin{footnotesize}
            \hrule\vspace{2pt}
            \nodeSet: Set of all nodes, part of the topology \\[5pt]
            \edgeSet: Set of all links, part of the topology \\[5pt]
            \commoditySet: Set of all traffic classes \\[5pt]
            \pathSet{\commodity}: Paths available for class 
            $\commodity \in \commoditySet$; output by path-selection 
            stage (\secref{sec:path})
            \vspace{2pt}\hrule
        \end{footnotesize}
    \end{minipage}
    \tightcaption{Network data input}
    \vspace{.1in}
    \label{fig:networkdata}
\end{figure}

\mypara{Data inputs}
There are two basic data inputs that the developer needs to provide to
any network optimization.  First, the network topology is a required
input, specified as a graph with \nodeSet and \edgeSet. It also
contains metadata of node/edge types or properties; e.g., nodes can
have designated functions like ``switch'' or ``middlebox''. Second,
\Name needs a specification of \emph{traffic classes}, where each
class \commodity has associated ingress and egress nodes and some
expected traffic volume.
Each class can (optionally) be associated with a specification of the
``processing'' required for traffic in this class, e.g., service
chaining.  Finally, to each traffic class \commodity is associated
a set \pathSet{\commodity} available to route flows in class \commodity;
\pathSet{\commodity} is output by a path-selection preprocessing step
described in \secref{sec:path}.

\newcolumntype{V}{m{.2em}}
\newcolumntype{P}{m{4.5cm}}
\begin{figure}[t]
    \begin{small}
    \begin{tabulary}{\columnwidth}{Vp{1cm}p{1.1cm}P}
\toprule
        & Var.\ & API & Description\\
    \midrule
        \multirow{6}{*}{\rotatebox[origin=c]{90}{\centering \textit{Decision}}} & 
         \flowFrac{\commodity}{\Path} & \xp(\commodity, \Path) & 
            Fraction of class-\commodity flows allocated to path $\Path \in
            \pathSet{\commodity}$; non-integer \\
	&\enabled{\Path} & \bp(\Path) & Is path \Path  used; binary \\
        &\enabled{\node} & \bn(\node) & Is node \node  used; binary \\
        &\enabled{\edge} & \be(\edge) & Is link \edge  used; binary \\
        & \nodeCapacityVar{\node}{\nodeResource} & \nc(\node, \nodeResource) &
            Capacity allocated for resource \nodeResource at node \node; non-integer\\
        \midrule
        \multirow{5}{*}{\rotatebox[origin=c]{90}{\centering \textit{Derived}}} &
		 \allocation{\commodity} & \al(\commodity) & 
            Fraction of \commodity's ``demand'' routed; non-integer \\ 
        & \edgeLoad{\edge}{\edgeResource} & \el(\edge, \edgeResource) &
            Amount of resource
            \edgeResource consumed by flows
            routed over link \edge; non-integer \\
        & \nodeLoad{\node}{\nodeResource} & \nl(\node, \nodeResource) &
            Amount of resource \nodeResource 
           consumed by flows routed via node \node; non-integer\\
        \bottomrule
    \end{tabulary}
    \end{small}
    \caption{Variables internal to the optimization}
    \label{fig:variables}
\end{figure}

\mypara{Internal variables}
\Name internally defines a set of variables summarized in
\figref{fig:variables}.  We reiterate that the developer does not need
to reason about these variables and uses a high-level mental model as
discussed earlier.\footnote{We also expose low-level APIs
  (\secref{sec:internals:low_level}) for advanced users.}  There are
two main kinds of variables:

\begin{packeditemize}

\item {\em Decision variables} that identify key optimization control
  decisions. The most fundamental decision variable is
  \flowFrac{\commodity}{\Path}, which captures traffic routing
  decisions and denotes the fraction of flow for a traffic class
  \commodity that path $\Path \in \pathSet{\commodity}$ carries. This
  variable is central to various types of resource management
  applications as we will see later.  To capture topological
  requirements (e.g., \secref{sec2:topology}), we introduce three
  binary decision variables \enabled{\Path}, \enabled{\node}, and
  \enabled{\edge} that denote whether each path, node or link
  (respectively) is enabled ($=1$) or disabled ($=0$).  The variable
  \nodeCapacityVar{\node}{\nodeResource} is the \FrName-given
  allocation of resource-\nodeResource to node \node.

\item {\em Derived variables} are functions defined over the above
  decision variables that serve as convenient ``shorthands''.
  \allocation{\commodity} denotes the total fraction of flow for class
  \commodity that is carried by all paths. The load variables
  \nodeLoad{\node}{\nodeResource} and \edgeLoad{\edge}{\edgeResource}
  model the consumption of resource \nodeResource on node \node and
  link \edge, respectively.  
\end{packeditemize}
\noindent
The (low-level) API calls in \figref{fig:variables} return the name of
the corresponding variable, which can be used to access its value in a
public map of variable names to values.

\newcolumntype{H}{>{\raggedright\arraybackslash}p{.4175\textwidth}}
\begin{figure*}
{\small
  \begin{tabulary}{\textwidth}{p{.1\textwidth}HH}
    \toprule
    Group & Function & Description\\
    \midrule
    \multirow{3}{*}{\parbox{7.5em}{\textit{Routing}\\($\commoditySubset \subseteq \commoditySet$)}} & \labelAllocateFlow & 
        Allocate flow in the network\\
    & \labelRouteAllFlow & Route all traffic demands\\
    & \labelOnePath(\commoditySubset)
        & For each $\commodity \in \commoditySubset$, at most one
        $\Path \in \pathSet{\commodity}$ is enabled.\\
    \hline
    \multirow{8}{*}{\textit{Capacities}}
    & \labelLinkCon(\edgeResource, \edgeCapacity, \edgeCapFunc)
        & If \edge is in \edgeCapacity, then limit utilization of
        link resource \resource on link \edge to $\edgeCapacity[\edge]$. \\
    & \labelNodeCon(\nodeResource, \nodeCapacity, \nodeCapFunc) 
        & If \node is in \nodeCapacity, then limit utilization of
        node resource \resource on node \node to $\nodeCapacity[\node]$. \\
    & \labelNodeConPerPath(\nodeResource, \nodeCapacity, \nodeCapFunc)
        & If \node is in \nodeCapacity, then limit utilization of
        node resource \resource on node \node by enabled paths to
        $\nodeCapacity[\node]$. \\
    & \labelCapacityBudget(\nodeResource, \nodeSubset, \totalCapacity)
        & Limit total type-\resource resources allocated to nodes in
        $\nodeSubset \subseteq \nodeSet$ to \totalCapacity.  Used when
        \Name is allocating capacities.\\
    \hline
    \multirow{10}{*}{\parbox{7.5em}{\textit{Topology\\control\\($\commoditySubset \subseteq \commoditySet$)}}} &
        \labelAllNodes(\commoditySubset) 
        & For each $\commodity \in \commoditySubset$ and each $\Path \in
        \pathSet{\commodity}$, \Path can be enabled iff
        all nodes on \Path are enabled.\\
    & \labelSomeNodes(\commoditySubset) 
        & For each $\commodity \in \commoditySubset$ and each $\Path \in
        \pathSet{\commodity}$, \Path can be enabled iff
        some node on \Path is enabled.\\
    & \labelAllEdges(\commoditySubset) 
        & For each $\commodity \in \commoditySubset$ and each $\Path \in
        \pathSet{\commodity}$, \Path can be enabled iff
        all links on \Path are enabled.\\
    & \labelEnforcePathDisable(\commoditySubset) 
        & For each $\commodity \in \commoditySubset$ and each $\Path \in
        \pathSet{\commodity}$, \Path can carry traffic only if it is
        enabled.\\
    & \labelNodeBudget(\nodeBudgetFunc, \nodeBudget)
        & Total cost of enabled nodes, as computed using \nodeBudgetFunc,
        is at most \nodeBudget.\\
    \hline
    \textit{Objective} & 
        \labelSetPredefObj(\name) & 
        Set one of the predefined functions as the objective 
        (see \figref{fig:objectives-high}).\\
    \bottomrule
\end{tabulary}
}
\tightcaption{Selected constraint template functions for building optimizations; see \figref{fig:funcs} for \edgeCapFunc, \nodeCapFunc, and \nodeBudgetFunc}
\label{fig:constraints}
\end{figure*}

Given these preliminaries, next we describe the key building blocks of a
\Name-based network optimization program. 

\subsection{Routing requirements}

Routing constraints control the allocation of flow in the network.
\labelAllocateFlow creates the necessary structure for routing the
traffic though a set of paths for each traffic class.  Some network
applications try to satisfy as much of their flow demands as possible
(e.g., max-flow) while others (e.g., TE) want to ``saturate'' demands.
For example, a developer of a TE application (\secref{sec2:te}) would like to
route all traffic though the network, and thus she would add the
following high-level routing constraint templates to her empty
\optObj:
\begin{lstlisting}[numbers=none, firstnumber=auto, name=TEexample]
(*\optObj.\labelAllocateFlow{}() *)
(*\optObj.\labelRouteAllFlow{}() *)
\end{lstlisting}
A developer writing a simple max-flow, however, would only need
\labelAllocateFlow since there is no requirement on saturating demands
in that case.

The $\labelOnePath(\commoditySubset)$ constraint forces a single
flow-carrying path per class $\commodity \in \commoditySubset$,
preventing flow-splitting and multipath routing.

\mypara{Internals} 
Formally, $\labelAllocateFlow$ ensures that the
total traffic flow across all chosen paths for the class $\commodity$
 matches  the variable $\allocation{\commodity}$. 
\[\displaystyle \forall \commodity \in \commoditySet :\;
\sum_{\Path \in \pathSet{\commodity}} \flowFrac{\commodity}{\Path} 
= \allocation{\commodity}\]
\noindent
Similarly,  \labelRouteAllFlow implies: 
\[≈\displaystyle \forall \commodity \in \commoditySet :\;
\allocation{\commodity} = 1\]
Due to space limitations, we do not provide the formal basis for
\labelOnePath.

\subsection{Resource capacity constraints}
As we saw in \secref{sec:background}, SDN optimizations have to deal
with a variety of capacity constraints dealing with network resources,
such as link bandwidth, switch rule capacities, middlebox CPU and
memory capacities.  \Name provides users the flexibility to write
custom resource management logic within a given network structure. For
this purpose we create several types of functions, depicted in
\figref{fig:funcs}. These functions allow the user to compute the
``cost'' of routing traffic through a link, a node, or a given
path. \Name provides multiple implementations of these for common
tasks, but allows the user to specify their own logic, as well, as we
will show later (\secref{sec:examples}).

For example, \labelLinkCon deals with constraints on the network
links, and \labelNodeCon incorporates constraints on the nodes.  For
our example, let us add a constraint that limits link usage.  For
this, we need a resource that we are constraining (in this case,
\emph{bandwidth}), a mapping of links to their
capacities,\footnote{When capacities are unknown a priori and should
  be allocated by the optimization itself, a capacity of \None
  (meaning To Be Allocated) can be specified, instead.} and
optionally, a way to compute the cost of traffic on a link.

\begin{lstlisting}[numbers=none, firstnumber=auto, name=TEexample]
(*\optObj.\labelLinkCon*)('bandwidth', 
    {(1,2): 10**7, (2,3): 10**7}, 
    defaultLinkFunction)
\end{lstlisting}
\noindent
This indicates that bandwidth should not exceed 10 Mbps for links 1-2
and 2-3. Note that the default function is purely for illustration;
the developer can write her own $\edgeCapFunc$ (recall
\figref{fig:funcs}).

\labelNodeConPerPath generates constraints on the nodes that do not
depend on the traffic, but rather on the routing path. That is, the
cost of routing at a node does not depend on the volume or type of
traffic being routed; it depends on the path and its properties. The
best example of such usage is accounting for the limited rule space on
a network switch (e.g., \secref{sec2:sc}). If a path is ``active'', the
rule must be installed on each switch to support the path.

\begin{figure}
    \begin{minipage}[t]{\columnwidth}
        \raggedright
        \begin{footnotesize}
        \hrule\vspace{2pt}
        $\edgeCapFunc(\edge,\commodity,\Path,\edgeResource)$:
        Amount of resource type \edgeResource consumed if all class-\commodity traffic is allocated to path $\Path \ni \edge$ for link \edge\\[5pt]
        $\nodeCapFunc(\node,\commodity,\Path,\nodeResource)$:
        Amount of resource \nodeResource consumed if all class-\commodity traffic is allocated to path $\Path \ni \node$ for node \node\\[5pt]
        $\nodeBudgetFunc(\node)$:
        Cost of using node \node; required with \labelNodeBudget \\[5pt]
        $\routingCostFunc(\Path)$:
        Cost of routing along path \Path; required with \labelRoutingCost \\[5pt]
        $\pathPredicate(\Path)$:
        Determine whether any given path is valid by returning \true or \false 
        \vspace{2pt}\hrule
        \end{footnotesize}
    \end{minipage}
\tightcaption{Customizable functions}
\label{fig:funcs}
\end{figure}

\mypara{Internals}
\labelLinkCon and \labelNodeCon rely on \edgeCapFunc and \nodeCapFunc,
respectively, to compute the cost of using a particular resource at a
link or node if all of the class-\commodity traffic was routed to
it. Internally, the load is multiplied by the
\flowFrac{\commodity}{\Path} variable to capture the load accurately,
then the load is capped by a user-provided \edgeCapacity
(\nodeCapacity), which is a mapping of links (nodes) to capacities for
a given \resource. (Similar node capacity equations not shown for
brevity.)
\begin{align*}
& \forall\edge \text{~in~} \edgeCapacity: \\
& \hspace{1.5em}\displaystyle \edgeLoad{\edge}{\edgeResource} =
\sum_{\commodity} \sum_{\Path \in \pathSet{\commodity}: \edge \in \Path} \!\!\!\!\!
\flowFrac{\commodity}{\Path} \times 
\edgeCapFunc(\edge,\commodity,\Path,\edgeResource)\\
& \hspace{1.5em}\displaystyle \edgeLoad{\edge}{\edgeResource} \le \edgeCapacity[\edge]
\end{align*}

The \labelNodeConPerPath functions a bit differently, as it depends on
enabled paths:
\begin{align*}
&  \displaystyle \forall \node \text{~in~} \nodeCapacity:\\
& \hspace{1.5em}\displaystyle \nodeLoad{\node}{\nodeResource} =
\sum_{\commodity} \sum_{\Path \in \pathSet{\commodity}: \node \in \Path} \!\!\!\!\!
\enabled{\Path} \times
\nodeCapFunc(\node,\commodity,\Path,\nodeResource) \\
& \hspace{1.5em}\nodeLoad{\node}{\nodeResource} \le  \nodeCapacityVar{\node}{\nodeResource} \\
& \hspace{1.5em}\text{if~} \nodeCapacity[\node] \neq \None \text{~then~} \nodeCapacityVar{\node}{\nodeResource} = \nodeCapacity[\node]
\end{align*}

\subsection{Node/link activation constraints}
These constraints, when used, allow developers to logically model the
act of \textit{enabling} or \textit{disabling} nodes, links, and
paths, e.g., for managing energy or other costs (e.g.,
\secref{sec2:topology}).  We identify two possible modes of
interactions between these topology modifiers and the optimization.
The optimization developer can choose the one that is most suitable
for their context: \labelAllNodes captures the property that disabling
a node disables all paths that traverse it; and \labelSomeNodes
captures the property that enabling a node permits any path traversing
it to be enabled, as well.  The latter version is suitable when, e.g.,
a node can still route traffic even if its other (middlebox)
functionality is disabled, and so a path containing that node is
potentially useful as providing middlebox functions if at least one of
its nodes is enabled.  Naturally there are analogous constraint
templates for links, but we omit them here for brevity.  A third
constraint template, \labelEnforcePathDisable, restricts a path to
carry traffic only if it is enabled.

For example, a developer trying to implement the application from \secref{sec2:topology} 
 can model the requirements for shutting off datacenter nodes by 
 adding the  \labelAllNodes and \labelEnforcePathDisable templates:
\begin{lstlisting}[numbers=none]
(*\optObj.\labelAllNodes(trafficClasses)*)
(*\optObj.\labelEnforcePathDisable(trafficClasses)*)
\end{lstlisting}

Other efficiency considerations may enforce a budget on the number of
enabled nodes, to model constraints on total power consumption of
switches/middleboxes, cost and budget of installing/upgrading
particular switches, etc.  These are captured via the \labelNodeBudget
constraint.

\mypara{Internals}
Internally, these topology modification templates are achieved using
the binary variables we introduced earlier. Specifically, the above
requirements can be formalized as follows:

\begin{small}
\[\begin{array}{@{\extracolsep{-0.25em}}ll}
& \forall \Path \in \pathSet{\commodity}: \\
\labelAllNodes
& \hspace{1em}\forall \node \in \Path:\; \enabled{\Path} \le \enabled{\node} \\
\labelSomeNodes
& \hspace{1em}\enabled{\Path} \le \sum_{\node \in \Path} \enabled{\node} \\
\labelEnforcePathDisable
& \hspace{1em}\flowFrac{\commodity}{\Path} \le \enabled{\Path}
\end{array}\]
\end{small}

\noindent
Naturally, similar constraints are constructed for links.  Note
that \labelEnforcePathDisable is crucial to the correctness of the
optimization in that it enforces that no traffic traverses a disabled
path.  For brevity, we do not provide the formal basis for
\labelNodeBudget.

\subsection{Specifying network objectives}
The goal of SDN applications is eventually to optimize some
network-wide objective, e.g., maximizing the network throughput,
load-balancing, or minimizing total traffic footprint.
\figref{fig:objectives-high} lists the most common objective
functions, drawing on the applications considered in
\secref{sec:background}.  For instance, the developer of a TE
application may want to implement the objective of minimizing the
maximum link load and thus add the following code snippet:

\begin{lstlisting}[numbers=none, firstnumber=auto, name=TEexample]
(* \optObj.\labelSetPredefObj(\labelMaxLinkLoad) *)
\end{lstlisting}

Other developers (e.g.,~\secref{sec2:nfv}) may want to minimize the total
routing cost and invoke a \labelRoutingCost objective.  This objective
is parameterized via $\routingCostFunc(\Path)$; i.e., developers can
plugin their own cost metrics such as number of hops or link weights.
As shown, we also provide a range of natural load-balancing templates.

\begin{figure}[t]
\begin{small}
\begin{tabulary}{\columnwidth}{lp{5cm}}
    \hline
    \labelTotalFlow
     & $\displaystyle \mbox{maximize~} \sum_{\commodity \in \commoditySet} \allocation{\commodity}$ \\
    \labelMaxNodeLoad(\resource)
      & $\displaystyle \mbox{minimize~} \max_{\node \in \nodeSet}~ \nodeLoad{\node}{\nodeResource}$ \\
    \labelMaxLinkLoad(\resource)
      & $\displaystyle \mbox{minimize~} \max_{\edge \in \edgeSet}~ \edgeLoad{\edge}{\edgeResource}$\\
    \labelRoutingCost
    & $\displaystyle \sum_{\commodity,\Path}  \routingCostFunc(\Path) \times \flowFrac{\commodity}{\Path}$\\
\hline
\end{tabulary}
\end{small}
\tightcaption{Common objective functions}
\label{fig:objectives-high}
\end{figure}

\subsection{Advanced users and low-level interface}
\label{sec:internals:low_level}

As we will see in \secref{sec:examples}, the \Name API is general and
expressive enough to capture the diverse requirements of the broad
spectrum of applications discussed in \secref{sec:background}. That
said, as part of the \Name design we also expose a low-level API that
gives more control to the user by giving access to the \Name internal
variables.  Advanced users can use this API for further
customization, as we will see in \secref{sec:examples}.

For instance, each API call in \figref{fig:variables} enables the name
of the corresponding internal variables to be retrieved.  Similarly,
using the \defVar(\name, \coeffs, \lowerBound, \upperBound) function,
the user can create a new variable with name \name, specify numeric
lower and upper bounds (\lowerBound and \upperBound), and equate it to
a linear combination of any other existing variables as specified by
\coeffs, a map from variable names to numeric coefficients.  This is a
useful primitive when specifying complex objectives. Finally, \Name
also allows to set a custom objective function that is a linear
combination of any existing variables. This is done using the
\setObj(\coeffs, \optDirection) function call, which accepts a mapping
\coeffs of variable names to their coefficients.  The binary input
\optDirection indicates whether the objective should be minimized or
maximized.

\section{Path generation and selection}
\label{sec:path}

Given these constraint templates, the remaining question is how we
populate the path set \pathSet{\commodity} for each traffic class
\commodity to meet two requirements. First, each $\Path \in
\pathSet{\commodity}$ should satisfy the desired policy specification
for the class \commodity. Second, \pathSet{\commodity} should contain
paths for each class \commodity that make the formulation tractable
and yet yield near-optimal results.  We describe how we address each
concern next.

\mypara{Generation} First, to populate the paths, \FrName does an offline
enumeration of all simple (i.e., no loops) paths per class.\footnote{This is
to simplify the forwarding rules without resorting to tunneling or packet
tagging~\cite{simple}.} Given this set, we filter out the 
 paths that do no satisfy the  user-defined predicate \pathPredicate such that
$\pathPredicate(\Path)=$\true only if \Path is a ``valid'' path. (We 
 can generalize this to allow different predicates per class; not shown 
 for brevity.)

In practice, we implement the predicate as a flexible \Python
callable function rather than constrain ourselves to specific notions
of path validity (e.g., regular expressions as in prior work~\cite{merlin}).
Using this predicate gives the user flexibility to capture a range of
possible requirements.  Examples include waypoint enforcement (forcing
traffic through a series of middleboxes in order); enforcing redundant
processing (e.g., through multiple NIDS, in case one fails open); and
limiting network latency by mandating shorter paths.

\mypara{Selection} Using all valid paths per
 class may be inefficient as the number of paths grows exponentially with the size of the network; i.e., the  LP/ILP \Name generates 
 will have an exponential number of rows and columns making it difficult to
solve in reasonable time.  \Name provides  path selection
algorithms that choose a subset of valid paths that are still likely
to yield near-optimal results in practice.  Specifically, two natural 
 methods work well across the spectrum  of applications  we 
 have considered: (1) \pathPruneNmbr shortest paths for
latency-sensitive applications (\pathPruneStrategy =
\pathPruneStrategyShortest) or (2) \pathPruneNmbr random paths for
applications involving load balancing (\pathPruneStrategy =
\pathPruneStrategyRandom).  \Name is flexible to incorporate other 
 selection strategies; e.g., picking paths with minimal node overlap for 
 fault tolerance.

\mypara{Developer API}
The developer  can specify the path predicate and selection strategy when
starting the optimization. Note that the developer does not 
 need to be involved in the low-level details of generation and pruning;  \Name
 runs these steps automatically.  We also provide APIs for developers to 
 add their own logic for generation and selection; we do not discuss 
 these due to space limitations.

\section{Examples}
\label{sec:examples}

Next, we show end-to-end examples to highlight the ease of using the  \Name APIs
to write existing and novel SDN network optimizations.  These examples
are actual \Python code that can be run, not just pseudocode (we refer the
reader to the anonymized manual for full code examples~\cite{manual}). By
comparison, the code is significantly higher-level and more readable than
the equivalent \CPLEX code would be, as it does not need to deal with
large numbers of underlying variables and constraints.

\begin{figure}[t]
\lstset{basicstyle=\scriptsize\ttfamily}
\begin{lstlisting}
SIMPLE_predicate = functools.partial(waypointMboxPredicate, order=('fw','ids')) (*\label{lst:simple:predicate}*)
def SIMPLE_NodeCapFunc(node, tc, path, resource, nodeCaps): (*\label{lst:simple:capfuncbegin}*)
    if resource == 'cpu' and node in nodeCaps['cpu']:
        return tc.volFlows * tc.cpuCost / nodeCaps[resource][node] (*\label{lst:simple:capfuncend}*)
capFunc = functools.partial(SIMPLE_NodeCapFunc, nodeCaps=nodeCaps) (*\label{lst:simple:curry}*)

def SIMPLE_TCAMFunc(node, tc, path, resource): (*\label{lst:simple:capfunctcam}*)
    return 1
# Path generation, typically run once in a precomputation phase
opt = getOptimization() (*\label{lst:simple:opt}*)
pptc = generatePathsPerTrafficClass(topo, trafficClasses, SIMPLE_predicate, 10, 1000, (*\hfill*) functools.partial(useMboxModifier, chainLength=2)) (*\label{lst:simple:pathgen}*)
# Allocate traffic to paths
pptc = chooserand(pptc, 5) (*\label{lst:simple:pathselect}*)
opt.addDecisionVariables(pptc) (*\label{lst:simple:constraintBegin}*)
opt.addBinaryVariables(pptc, topo, ['path', 'node'])
opt.addAllocateFlowConstraint(pptc)
opt.addRouteAllConstraint(pptc)
opt.addLinkCapacityConstraint(pptc, 'bandwidth', linkCaps, defaultLinkFuncNoNormalize)
opt.addNodeCapacityConstraint(pptc, 'cpu', (*\hfill*) {node: 1 for node in topo.nodes() if 'fw' or 'ids' in topo.getServiceTypes(node)}, capFunc) (*\label{lst:simple:normcap}*)
opt.addNodeCapacityPerPathConstraint(pptc, 'tcam', nodeCaps['tcam'], SIMPLE_TCAMFunc) (*\label{lst:simple:constraintEnd}*)
opt.setPredefinedObjective('minmaxnodeload', 'cpu') (*\label{lst:simple:objective}*)
opt.solve() (*\label{lst:simple:solve}*)
obj = opt.getSolvedObjective() 
pathFractions = opt.getPathFractions(pptc)
c = controller()
c.pushRoutes(c.getRoutes(pathFractions)) (*\label{lst:simple:push}*)
\end{lstlisting}
\vspace{-0.3cm}
\tightcaption{\Python code to express SIMPLE~\cite{simple} in \Name}
\label{fig:simple}
\vspace{-0.3cm}
\end{figure}

\mypara{Service chaining (\secref{sec2:sc})} As a concrete instance of the
service chaining example, we consider SIMPLE~\cite{simple}.
SIMPLE involves the following requirements: route all traffic through the
network, enforce the service chain (e.g., ``firewall followed by IDS'') policy
for all traffic, load balance across middleboxes, and do so while respecting
CPU, TCAM, and bandwidth requirements.
\figref{fig:simple} shows how SIMPLE optimization can be written in $\approx$ 25
lines of code.  This listing assumes that topology and traffic classes have been
set up, in the \texttt{topo} and \texttt{trafficClasses} objects, respectively.

The first part of the figure shows function definitions and the path generation
step, which would typically be performed once as a precomputation step.  We
start by defining a path predicate (line~\ref{lst:simple:predicate}) for basic
enforcement through middleboxes by using the \Name-provided function with the
middlebox order.
The next few lines
(lines~\ref{lst:simple:capfuncbegin}-\ref{lst:simple:capfuncend}) show  a
custom node capacity function to normalize the CPU load between 0 and 1.  This
computes the processing cost per traffic class (number of flows times CPU cost)
normalized by the current node's capacity.
Similarly,  the TCAM capacity function
captures  that each path consumes a 
single rule per switch (line~\ref{lst:simple:capfunctcam}). 
The user gets the
optimization object (line~\ref{lst:simple:opt}), and generates the
paths (line~\ref{lst:simple:pathgen}), obtaining the ``paths per traffic class''
(\texttt{pptc}) object. The path generation algorithm is parameterized with the
custom \texttt{SIMPLE\_predicate}, a limit on path length of 10 nodes, and a
limit on the number of paths per class of 1000.
It is also instructed to evaluate every possible use of two
middleboxes on a routing path for inclusion as a distinct path in the output.

The remaining lines show what would be executed whenever a new
allocation of traffic to paths is
desired. Line~\ref{lst:simple:pathselect} selects 5 random paths per
traffic class;
lines~\ref{lst:simple:constraintBegin}--\ref{lst:simple:constraintEnd}
add the routing and capacity constraints.  We use the default link
capacity function for bandwidth constraints, and our own functions for
CPU and TCAM capacity. Because the CPU capacity function normalizes
the load, the capacity of each node is now 1
(line~\ref{lst:simple:normcap}).  The user then selects a predefined
objective to minimize the CPU load (line~\ref{lst:simple:objective})
and calls the solver (line \ref{lst:simple:solve}). Finally, the
program gets the results and interacts with the SDN controller
interface to automatically install the rules
(line~\ref{lst:simple:push}).

\mypara{ElasticTree~\cite{elastictree}}
Due to space limitations we only show the most important difference
between ElasticTree and SIMPLE. First, there is no requirement on
paths, and so \texttt{nullPredicate} is used for path
generation. Also, we use link binary variables (see
line~\ref{lst:elastic:bin} below), as well as the node/link activation
constraints
(lines~\ref{lst:elastic:toggleBegin}--\ref{lst:elastic:toggleEnd}).
Finally, we must use the low-level API to define power consumption for
switches and links (lines~\ref{lst:elastic:switchpower},
\ref{lst:elastic:linkpower}) and use these variables to define a custom objective
function (line~\ref{lst:elastic:objective}).

\lstset{basicstyle=\scriptsize\ttfamily}
\begin{lstlisting}
opt.addBinaryVariables(pptc, topo, ['path', 'node', 'edge']) (* \label{lst:elastic:bin} *)
opt.addRequireAllNodesConstraint(pptc) (* \label{lst:elastic:toggleBegin} *)
opt.addRequireAllEdgesConstraint(pptc)
opt.addPathDisableConstraint(pptc) (* \label{lst:elastic:toggleEnd} *)
opt.defineVar('SwitchPower', {opt.bn(node): switchPower[node] for node in topo.nodes()}) (* \label{lst:elastic:switchpower} *)
opt.defineVar('LinkPower', {opt.be(u, v): linkPower[(u, v)] for u, v in topo.links()}) (* \label{lst:elastic:linkpower} *)
opt.setObjectiveCoeff({'SwitchPower': .75, 'LinkPower': .25}, 'min') (* \label{lst:elastic:objective} *)
\end{lstlisting}

\mypara{New elastic scaling capabilities}
Finally, we show  \Name can be used for novel SDN applications. Specifically, 
we consider a elastic NFV setting~\cite{stratos} that 
places middleboxes in the network and allocates
capacities  on demand in response to observed demand. 
There could be additional constraints, such
as the total number of such VM locations.
 As a simple objective, we consider an upper bound on the number of nodes used 
   while still  load balancing across  virtual middlebox instances. 
  We can easily add other objectives such as minimizing number of VMs.
For brevity we highlight only the key parts of building
such a novel application. 

\begin{lstlisting}
predicate = hasMboxPredicate (* \label{lst:scaling:predicate} *)
opt.addBinaryVariables(pptc, topo, ['path', 'node'])
opt.addNodeCapacityConstraint(pptc, 'cpu', {node: 'TBA' for node in topo.nodes()}, lambda node, tc, path, resource: tc.volFlows * tc.cpuCost) (* \label{lst:scaling:capacity} *)
opt.addRequireSomeNodesConstraint(pptc) (* \label{lst:scaling:somenodes} *)
opt.addPathDisableConstraint(pptc) (* \label{lst:scaling:pathdis} *)
opt.addBudgetConstraint(topo, lambda node: 1, topo.getNumNodes()/2) (* \label{lst:scaling:budget} *)
opt.setPredefinedObjective('minmaxnodeload', resource='cpu') (* \label{lst:scaling:objective} *)
\end{lstlisting}

First, we define a valid path to be one that goes though a middlebox; \Name
provides a predicate for that (line~\ref{lst:scaling:predicate}). The main
difference here is the definition of capacities with the \None value on line
\ref{lst:scaling:capacity}; this indicates that our optimization  must allocate
the capacities to the nodes.  (\Name ensures that disabled nodes have  0
capacity  allocated.)  Thus we require at least one enabled node per path
(lines~\ref{lst:scaling:somenodes}, \ref{lst:scaling:pathdis}), limit the
number of enabled nodes (line~\ref{lst:scaling:budget}), and set the objective
(line~\ref{lst:scaling:objective}).

\section{Implementation}
\label{sec:implementation}

Next, we briefly describe how we prototype the various 
aspects of \Name.

\subsection{Core functionality}

\mypara{Developer interface} We currently provide a \Python API for SDN
optimization that is an extended version of the interface described in
\secref{sec:interface}.

\mypara{Solvers} We use \CPLEX as our underlying solver and use the
\Python API that \CPLEX already supports. The choice largely reflects our
familiarity with the tool, and we can substitute \CPLEX with other solvers like
\Gurobi, which has similar capabilities. \Name also has APIs to exploit solver
capabilities to use a previously computed solution and incrementally find a new
solution.  This approach is typically faster than starting from scratch and
so is useful for faster reconfigurations.

\mypara{Path enumeration}
Path generation is an inherently parallelizable process; we simply
launch separate \Python processes for different traffic classes.  We
currently support two path selection algorithms:
\pathPruneStrategyRandom and \pathPruneStrategyShortest.  It is easy
to add more algorithms as new applications emerge.

\mypara{Rule generation and control interface}
We use \OpenDaylight as the SDN control platform and use its REST API
to install the relevant rules.  We generate the rules based on the
optimization output, using network prefix splitting to implement the
fractional responsibilities represented by the
\flowFrac{\commodity}{\Path} variables.  This step is similar to prior
works that have mapped fractional processing and forwarding
responsibilities to flows in the network (e.g.,~\cite{openflowlb,snips}),
and so we do not repeat it here.

\subsection{Extensions}
We have implemented additional features in \Name that 
 are  useful for responding to traffic changes or other events.

\mypara{Minimizing reconfiguration changes} Networks are in flux
during reconfigurations with potential performance or consistency
implications, and thus we want to minimize the amount of path churn.
We extend the basic APIs to support a new constraint that bounds (or
minimizes) the logical distance between the previous solution and the
new solution to help minimize the number of flows that have to be
assigned a new route.  To this end, \Name also supports path selection
that gives priority to previously selected paths.

\mypara{Reacting to faults} We provide a basic mechanism to react to node or
link failures as follows.   We build a  dependency graph between the set of
paths we have chosen during generation (\secref{sec:path}) and the specific
links, switches, and middlebox nodes that these paths traverse.  Given this,
we implement the following two-step heuristic.  As a first step, we simply
re-run the optimization with the old set of selected paths but excluding the
specific paths impacted by the failures. If the objective value is much worse,
then we fall back to  running the selection step again (i.e., selecting
\pathPruneNmbr paths from the already generated paths for each traffic class,
either \pathPruneStrategyRandom or \pathPruneStrategyShortest) and rerun the
optimization.  This two-step heuristic works well in the common case because
(i) the time to run the optimization is  lower than the time for selection, and
(ii) we can often find  near-optimal solutions even with fewer paths per
 class (\figref{fig:runtimeVsNumPaths}).

\section{Evaluation}
\label{sec:eval}

In this section we show that \Name is practical and yields optimal
solutions for an array of representative applications and topologies,
at timescales that are often dramatically better than custom
solutions.  We specifically evaluate the effect of using \Name to
implement three existing SDN applications:
ElasticTree~\cite{elastictree}, SIMPLE~\cite{simple}, and
Panopticon~\cite{panopticon}.  For each application, we implemented
the original optimization formulation presented in prior
work. Specifically, this is an ILP in the case of ElasticTree, SIMPLE,
and Panopticon, and its solution (where we could compute it) is
denoted as an ``original solution'' below.  We also implemented in
\FrName the ``new elastic scaling'' application from
\secref{sec:examples}, to which we refer as ``ElasticScaling'' below.
We do not have a competing non-\FrName implementation for
ElasticScaling, however.  Finally, in order to draw comparisons to a
conceptually similar, recent effort (Merlin~\cite{merlin}) in
\secref{sec:eval:opt}, we implemented the example application used in
that paper, in \FrName.

Our evaluation has three parts.  In
\secref{sec:eval:opt}, we demonstrate that \FrName generates optimal
(or very nearly optimal) solutions, but does so orders-of-magnitude
faster than their original formulations.  In \secref{sec:eval:dev}, we
describe the improvements to developer effort (e.g., in lines of code
needed) that \FrName provided in our implementations.  In
\secref{sec:eval:sensitivity}, we evaluate the sensitivity of \FrName
solutions to its path selection parameters.

To perform these evaluations, we chose topologies of various sizes
from the TopologyZoo dataset~\cite{topologyzoo}.  For ElasticTree, we
also constructed FatTree topologies of various
sizes~\cite{al2008scalable}; we refer to these as ``kX'' where X
denotes the arity of the FatTree, as defined in prior work.  Lacking
traffic matrices for these networks, we generated them synthetically:
we used a uniform traffic matrix for the ``kX'' networks and a
gravity-based model~\cite{r:05} for the TopolozyZoo dataset.  Randomly
sampled values from the log-normal distribution served as
``populations'' for the gravity model. In our experience, the choice
of traffic matrix does not qualitatively impact the results we
present.  Unless otherwise specified, we use 5 paths per traffic class
when running \Name.  All times below refer to computation on computers
with 2.4GHz cores and 128GB of RAM.  In the interest of space, we do
not present all measures for all applications, but our reported
observations are representative across all applications and topologies
considered.

\subsection{Optimality and scalability}
\label{sec:eval:opt}

First, we examine how well \Name's results match original solutions,
which are themselves optimal (by definition). \tabref{tab:optratio}
shows the ``objective ratio'' of the \FrName solution's objective
value to the original solution's objective value, for topologies and
applications for which we were able to obtain original solutions.  A
ratio of 1 indicates that \Name obtains an optimal solution,
which it does in most of the cases.  ``N/A'' indicates that we could
not obtain an original solution within 30 minutes of computation.

\begin{table}
\renewcommand{\tabcolsep}{0.08cm}
\footnotesize
\begin{tabulary}{\columnwidth}{lcccccc}
{} &  Abilene &  Quest &  Geant2012 &  Bellcanada &  Dfn &  k4  \\
\midrule
ElasticTree &           N/A &         N/A &             N/A &              N/A &       N/A &      1.1 \\
Panopticon  &             1 &           1 &               1 &                1 &         1 &      N/A \\
SIMPLE      &             1 &           1 &             N/A &              N/A &       N/A &      N/A 
\end{tabulary}
\tightcaption{Objective ratio of \Name vs.\ original solutions}
\label{tab:optratio}
\end{table}

\begin{figure}
    \includegraphics{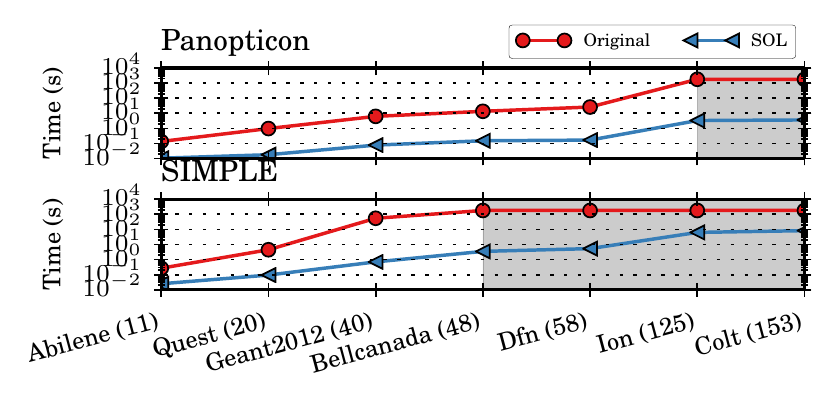}
    \tightcaption{Optimization runtime of \Name and the original
      formulations; gray regions indicate where original formulation
      could not be solved within 30 mins}
    \label{fig:runtimeVsOriginal}
\end{figure}

\Name solution times are at least one order of magnitude faster than
is solving the original formulations, and are often two or even three orders of
magnitude faster.  \figref{fig:runtimeVsOriginal} shows run times to find
original solutions.  Again, the runtime was capped at 30 min (1800 s), after
which the execution was aborted. Several original formulations did not complete
in that time, such as SIMPLE for topologies Bellcandada and larger, and
Panopticon for Ion and larger.  The topologies for which original solutions
could not be found are indicated in the gray regions in
\figref{fig:runtimeVsOriginal}.

\begin{figure}
\includegraphics{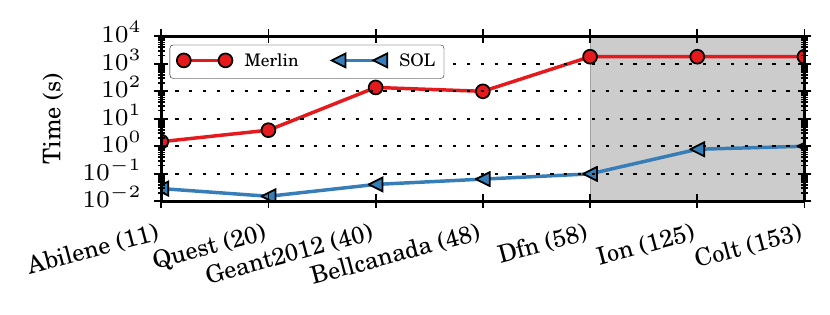}
\tightcaption{Comparison of optimization runtimes of \Name vs Merlin; gray
  region indicates where Merlin did not find solution within 30 mins}
\label{fig:runtimeVsMerlin}
\end{figure}

\mypara{Comparison to Merlin}
Merlin~\cite{merlin} is a recent work that tackles problems of network
resource management similar to \Name. While the goals and formulations
of Merlin and \Name are quite different, a comparison highlights the
generality of \Name and the power of its path abstraction.
Specifically, Merlin uses a more heavyweight optimization that is
always an ILP~\cite{merlin} and operates on a graph that is substantially
larger than the physical network. We implemented the example
application taken from the Merlin paper using both \Name and
Merlin. \figref{fig:runtimeVsMerlin} shows that \Name outperforms
Merlin by two or more orders of magnitude.

\mypara{Path selection and generation costs}
Path selection times are small, ranging from 0.1 to 3 seconds across
topologies.  Path selection is preceded by a path generation phase
that enumerates the simple paths per class.  Path generation is
moderately costly for large topologies, e.g., taking <300 s for the
largest presented topology, when parallelized to 60 threads.  However,
we highlight that path generation can be relegated to an offline
pre-computation phase that is only performed once.

\subsection{Developer benefits}
\label{sec:eval:dev}

We believe that \FrName is a much simpler framework for encoding SDN
optimization tasks, versus developing custom solutions by hand for
solution by an off-the-shelf solver.  In an effort to demonstrate this
simplicity somewhat quantitatively, \tabref{tbl:dev_effort}
shows the number of lines of \FrName code in our implementations for
the various applications (``\Name lines of code''), and the ratio of
the lines of code in our hand-developed scripts\footnote{Hand-coding
original formulations for direct consumption by \CPLEX would imply
repeating that process for each topology, and so we scripted the
generation of original formulations in \Python as efficiently as we
could.} to produce the original formulations to the lines of code
for our \FrName implementations (``Estimated improvement'').  We acknowledge
that lines-of-code comparisons are inexact, at best, but we do not
know of other ways of comparing ``development effort'' without
conducting user studies.  (We are considering this option for future
work.)

\begin{table}[th]
\footnotesize
\centering
\begin{tabulary}{\columnwidth}{lCC}
    Name & \Name lines of code & Estimated improvement\\
    \midrule
    ElasticTree & 16 & $21.8\times$\\
    Panopticon & 13 & $25.7\times$\\
    SIMPLE & 21 & $18.6\times$\\
    ElasticScaling & 15 & N/A
\end{tabulary}
\tightcaption{Development effort benefits provided by \Name}
\label{tbl:dev_effort}
\end{table}

There are reasons to believe, however, that the improvements indicated
in \tabref{tbl:dev_effort} are even conservative.  First, in our
opinion, our scripts for producing original formulations are
considerably more complex and delicate than our \FrName code.  We
primarily attribute this difference to needing to account for \CPLEX
particulars at all; with \FrName, these particulars are completely
hidden from the developer.  Second, \FrName translates
its optimization results to device configurations, whereas this
functionality is not even included in our scripts for producing
original formulations.  Producing device configurations from original
solutions would require designing an algorithm to map the control
variables in each formulation to relevant device configurations.

\subsection{Sensitivity}
\label{sec:eval:sensitivity}

\FrName solutions require the specification of both a number
(\pathPruneNmbr) and type (\pathPruneStrategyShortest or
\pathPruneStrategyRandom) of paths to select per traffic class.  In
this section we quantify how sensitive \Name is to these
path selection parameters.

\begin{figure}[th]
    \includegraphics[width=\columnwidth]{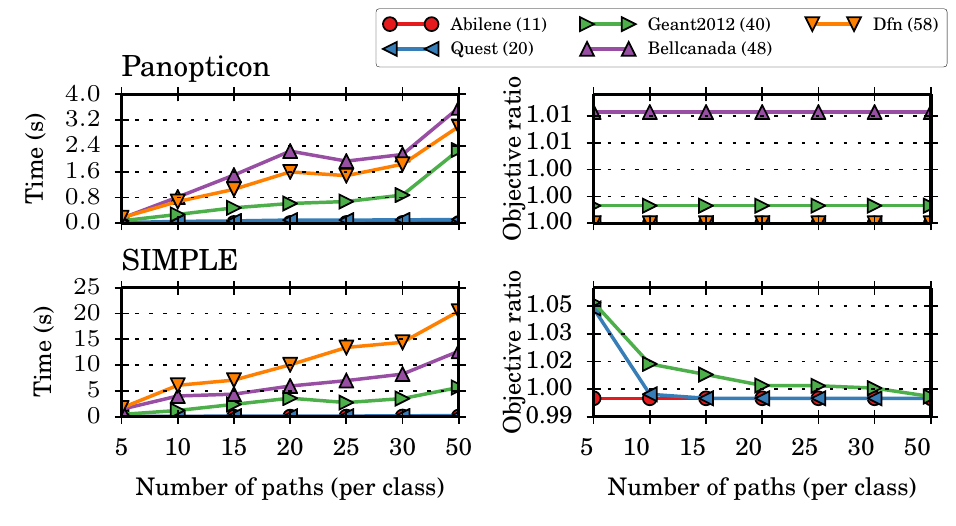}
    \tightcaption{Runtime and objective ratio as function of paths; optimality is
    achieved in most cases with as few as 5 paths per class}
    \label{fig:runtimeVsNumPaths}
\end{figure}

\mypara{Number of Paths}
\figref{fig:runtimeVsNumPaths} shows the runtime and the objective
ratio as a function of the number of paths per class for two
applications, SIMPLE and Panopticon. Unsurprisingly, with larger
number of paths, the runtime increases.  However, this is not a
significant concern, since we find optimal solutions at \pathPruneNmbr
as low as $5$.  These numbers are representative of all applications
and topologies we have considered.

\begin{figure}[th]
    \includegraphics[width=\columnwidth]{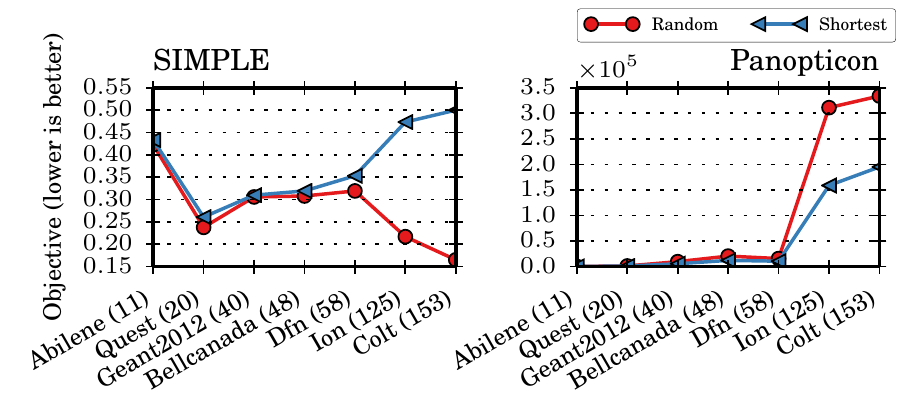}
    \tightcaption{\FrName objective values based on the chosen path
      selection strategy}
    \label{fig:objVsStrategy}
\end{figure}

\mypara{Path selection strategy}
\figref{fig:objVsStrategy} shows the difference in the
\FrName-computed objective function based on the different path
selection strategies as well as the original solution (where obtainable). This
figure indicates that choosing an appropriate path selection strategy can
provide substantial benefits for large topologies.  In our experience, most
problems lend themselves to a fairly obvious path selection strategy: those with
need for load balancing should use \pathPruneStrategyRandom and those that are
latency-sensitive should use \pathPruneStrategyShortest.  If in doubt, however,
both strategies can be attempted.

\section{Related Work}
\label{sec:related}

We  already discussed the optimization applications that  motivated \Name. Here
we focus on other related work.

\mypara{Higher-layer abstractions for SDN}    
This work includes new programming languages (e.g.,~\cite{pyretic,frenetic}),
testing and verification tools (e.g,~\cite{headerspaceanalysis}), semantics for
network updates (e.g.,~\cite{consistentupdate}), compilers for  rule generation
(e.g.,~\cite{onebigswitch}), abstractions for handling control conflicts (e.g.,
~\cite{corybantic}), and  APIs for users to express requirements
(e.g.,~\cite{pane}). These are orthogonal  and do not address
optimization in SDN applications, which is the focus of \Name.

 \mypara{Languages for optimization} There are several  modeling
frameworks such as {\tt AMPL}~\cite{ampl}, {\tt PyOpt}~\cite{pyopt}, and {\tt
PuLP}~\cite{pulp} for expressing optimization tasks.  However, these 
do not  specifically simplify network optimization.
\Name is a domain-specific library that operates at a higher level of
semantics than these ``wrappers''.  \Name offers a path-based
abstraction for writing network optimizations, exploits this structure
to solve these optimizations quickly, and generates network device
configurations that implement its solutions.

\mypara{Network resource management} Merlin is a language for network
resource management~\cite{merlin}.  In terms of the specific
applications that it can support, Merlin is restricted to using path
predicates expressed as regular expressions.  Our experiments suggest
that \Name is three orders of magnitude faster than Merlin using the
same underlying solvers.  That said, Merlin's ``language-based''
approach provides other capabilities (e.g., verified delegation) that
\Name does not (try to) offer.  Other works focus on traffic-steering
optimization (e.g.,~\cite{simple,cao2014traffic}).  \Name offers a
unifying abstraction that covers many network management applications.
\section{Conclusion}
\label{sec:conclusion}

Optimization is a core ingredient in designing many new SDN
applications.  Despite its broad utility, few efforts have attempted
to make optimization more accessible to potential SDN application
developers and network administrators.  Our vision is a general and
efficient high-level framework for expressing and solving complex
network optimization tasks. We showed that \Name achieves both
generality and efficiency via a path-centric optimization abstraction.
This abstraction provides the generality to capture diverse
applications, enables efficient solutions via simple path selection
algorithms, and also simplifies SDN rule generation.
We showed that \Name can concisely capture optimization applications
with diverse goals (traffic engineering, offloading, topology
modification, service chaining, etc.), that \Name yields optimal or
near-optimal solutions, and that \Name does so with far better online
performance than handcrafted optimization formulations.  We thus
believe that \Name can significantly lower the barrier of entry and
dramatically reduce development effort for novel SDN network
optimization applications.

\bibliographystyle{abbrv}
\bibliography{bibliography}

\end{document}